\documentclass[aps,superscriptaddress,showpacs,nofootinbib,eqsecnum,prd,notitlepage,twocolumn]{revtex4} 

\usepackage{hyperref}
\usepackage{eucal} 
\usepackage{amssymb}
\usepackage{amsmath} 
\usepackage{graphicx} 
\usepackage{epsfig}

\begin{document}

\title{Bivelocity picture in the
nonrelativistic limit of relativistic hydrodynamics}

\author{Tomoi Koide}  
\email{tomoikoide@gmail.com, koide@if.ufrj.br}
\affiliation{Instituto de F\'{\i}sica, Universidade Federal do Rio de Janeiro, 
C.P. 68528, 21945-970, Rio de Janeiro, Brazil}

\author{Rudnei O. Ramos} 
\email{rudnei@uerj.br} 
\affiliation{Departamento de F\'{\i}sica Te\'orica, Universidade do
  Estado do Rio de Janeiro, 20550-013 Rio de Janeiro, RJ, Brazil}

\author{Gustavo S. Vicente} 
\email{gsvicente@uerj.br, gustavosvicente@gmail.com}
\affiliation{Departamento de F\'{\i}sica Te\'orica, Universidade do
  Estado do Rio de Janeiro, 20550-013 Rio de Janeiro, RJ, Brazil}

\begin{abstract}

We discuss the nonrelativistic limit of the relativistic
Navier-Fourier-Stokes (NFS) theory.   The next-to-leading order
relativistic corrections to the NFS theory for the Landau-Lifshitz
fluid are obtained.  While the lowest order truncation of the velocity
expansion leads to the usual NFS equations of nonrelativistic fluids,
we show that  when the next-to-leading order relativistic corrections
are included,  the equations can be expressed concurrently with two
different fluid velocities. One of the fluid velocities is parallel
to the conserved charge current (which follows the Eckart definition)
and the other one is parallel to the energy current (which follows the
Landau-Lifshitz definition).  We compare this next-to-leading order
relativistic hydrodynamics with bivelocity hydrodynamics,  which is
one of the generalizations of the NFS theory and is formulated in such
a way to include the usual mass velocity and also a new velocity,
called the volume velocity. We find that the volume velocity can be
identified with the velocity obtained in the Landau-Lifshitz
definition.  Then, the structure of bivelocity hydrodynamics, which is
derived using various nontrivial assumptions, is reproduced in the
NFS theory including the next-to-leading order relativistic
corrections.  

\end{abstract}


\maketitle


\section{Introduction}
\label{intro}

Macroscopic matter, such as fluids, consists of extraordinary large
number of microscopic particles and the dynamics is determined by
solving highly coupled equations.  However, it is also known that the
long-wavelength and low-frequency behaviors are approximately
described by a coarse-grained dynamics called the
Navier-Fourier-Stokes (NFS) theory.  This coarse-grained dynamics is
much more tractable than the original microscopic dynamics, and has
been applied to various nonrelativistic collective phenomena
successfully.  Thus, it is quite natural to expect that this approach
is also useful for the application to relativistic phenomena.  As a
matter of fact, relativistic hydrodynamics has been used to study
relativistic collective behaviors in astrophysics, cosmology and
nuclear physics.

Despite the widespread use of relativistic hydrodynamic models, their
theoretical properties are still not fully understood because of the
difficulties inherent in the relativistic kinematics.  For example, it
is well-known that first-order dissipative relativistic theories have
problems concerning causality, generic stability and in general they
do not have a well-posed initial  value formulation (see, for example,
Hiscock and Lindblom \cite{hiscock}).  Other proposals for
relativistic hydrodynamics led to   dissipative relativistic theories
satisfying causality \footnote{However, the complexity of the generic
  stability of these theories led some authors to reconsider first
  order theories in order to establish this property
  properly~\cite{various}.} such as the second order theory by Israel
and Stewart \cite{israel,israelstewart}.

In this work we investigate another aspect of relativistic
hydrodynamics, that is, the nonuniqueness of the definitions of fluid
velocities.  {}For example, the conserved energy and charge densities
in relativistic systems are given by the sum and subtraction of the
particle and anti-particle contributions, respectively. Thus, the
flows of energy and charge are, in general, not parallel to each other
and we observe two different definitions for the fluid velocities.  In
one case we can define the fluid velocity to be parallel to the charge
current, and the other possibility is when the fluid velocity is
chosen to be parallel to the energy current.  The former case was
introduced by Eckart~\cite{eckart}  and the local rest frame
associated with this velocity is called the Eckart frame.  By
definition, there are no spatial components of the charge current in
the Eckart frame.  It should be noted, however, that  the Eckart
definition is not applicable, for example, in relativistic heavy-ion
collisions at vanishing baryon chemical potentials  and in early
universe cosmology, where the flow of a conserved charge is usually
not considered.  The other possibility for defining the fluid
velocity, and also of common usage, was proposed by
Landau-Lifshitz~\cite{landau}. In this case the fluid velocity is
chosen to be parallel to the energy current, i.e., there are no
spatial components of the energy current in this rest frame, called
the Landau-Lifshitz frame. 

As is well-known, in the nonrelativistic NFS theory, the Eckart and
Landau-Lifshitz rest frames are equivalent and conserved charge and
energy of fluids  are transported by an unique fluid velocity (the
mass velocity). However, when relativistic corrections are considered,
from the argument above, it is natural to expect that deviations from
the NSF theory  should depend on the choice of fluid velocity because
the definitions  of the local rest frame are changed.  The
next-to-leading order (NLO) corrections to the standard NSF fluid
equations were first considered by Chandrasekhar~\cite{chandrasekhar}
for ideal fluids. That study was later extended by
Greenberg~\cite{greenberg}  for nonideal fluids.  However, only the
Eckart definition of fluid velocity was considered  and the role of
the two fluid velocities in the modified hydrodynamics  was not
discussed.  Moreover, the two different local rest frames required in
the definitions of  nonrelativistic limit of hydrodynamic variables,
like the energy density and the conserved charge density,  were not
distinguished.  We extend those earlier works and apply to the case of
Landau-Lifshitz fluids,  determining the NLO relativistic corrections
to the NSF theory for the Landau-Lifshitz definition of  fluid
velocity in this paper. 

Then it is interesting to  contrast the obtained NLO relativistic
equations with a  recent proposal of the modification of the NFS
theory by
Brenner~\cite{brenner,brenner2,brenner2012,brenner2012-2,brenner2013}
to clarify the role of the two fluid velocities.  Since the velocity
of a tracer particle of nonrelativistic fluids is not necessarily
parallel to the mass velocity, it is claimed in his bivelocity
formulation that the existence  of these two velocities should be
included in a consistent formulation of the nonrelativistic
hydrodynamics.  This additional fluid velocity is called volume
velocity.  So far, there are various studies following this scenario
(for a list of related works, although far from complete, see, e.g.,
Refs.~\cite{dzy,kli,ott,grau,gree,daz,eu,don,koide}), but  it is still
controversial whether this bivelocity scenario is realized or not.

As mentioned, as far as the presence of the two definitions of
velocities is concerned,  the bivelocity argument is similar to what
is familiar in the community of relativistic hydrodynamics.  Then it
is interesting to discuss the structure of the NLO relativistic
corrections from the point of view  of bivelocity hydrodynamics,
because, as will be discussed in this paper, the corrections stem
from the fact that the nonrelativistic energy density is defined in
the Landau-Lifshitz frame while  the nonrelativistic conserved charge
density is in the Eckart frame.  In other words, the NLO relativistic
corrections are directly affected by the difference  of the two fluid
velocities.  Therefore, the detailed analysis of the NLO corrections
is useful even to inspect  the consistency of the structure of
bivelocity hydrodynamics.

In the present work we also study the formulation of bivelocity
hydrodynamics by comparing it to relativistic
hydrodynamics \footnote{It is well-known that  higher order kinetic
  corrections to the  NFS theory leads to the  Burnett and
  super-Burnett equations~\cite{colin}.  The relation between these
  kinetic corrections and the bivelocity picture was  discussed in
  Refs.~\cite{brenner2012,brenner2012-2}.}.    We start by considering
the nonrelativistic limit of relativistic hydrodynamics in the
Laudau-Lifshitz frame, and  we show that the standard NFS theory is
reproduced in the leading order approximation.  Moreover, it is found
that the derived hydrodynamics at the NLO can be cast into a form of
bivelocity hydrodynamics which is generalized so as to permit to
include the effect of relativistic corrections.

The rest of this paper is organized as follows.  In Sec.~\ref{sec2},
we briefly summarize relativistic hydrodynamics and discuss the
different  possibilities of the choice of fluid velocities in the
context of both Landau-Lifshitz and Eckart theories. In
Sec.~\ref{sec:leading_order},  we express the nonrelativistic limit of
the various hydrodynamic variables in the relativistic theory in terms
of the corresponding nonrelativistic ones. The leading order
truncation of the velocity expansion is implemented and we derive the
NFS theory. In Sec.~\ref{sec4}, we discuss the NLO corrections. Our
results, including the NLO corrections,  are then contrasted with
bivelocity hydrodynamics in Sec.~\ref{sec6}. Section~\ref{sec7} is
devoted to the concluding remarks.


\section{Relativistic Hydrodynamics}
\label{sec2}

\subsection{Ideal fluid}

In relativistic hydrodynamics, the energy-momentum tensor and the
conserved charge current are expressed in terms of hydrodynamic
variables describing the macroscopic motion of many-body systems
\footnote{The time scale of the evolution of  non-conserved quantities
  are  considered to be short and  these are usually not included as
  hydrodynamic variables.}.  In the case of an ideal fluid, two proper
scalar densities  ($\varepsilon$ and $P$) and one four-vector field
(the four-velocity $u^\mu$) are used to express the energy-momentum
tensor,

\begin{equation}
T_0^{\mu\nu} = (\varepsilon + P)u^{\mu} u^{\nu} - g^{\mu\nu}P\;,
\label{T0munu}
\end{equation}
where the Lorentz four-velocity field $u^{\mu}$ is expressed as 

\begin{eqnarray}
u^{\mu} &=& \gamma (1,{\bf v}/c),
\end{eqnarray}
and $\gamma = 1/\sqrt{1-{\bf v}^2/c^2}$ is the usual Lorentz factor,
with the spatial velocity ${\bf v}$ and the speed of light $c$.  The
four-velocity is normalized such that  $u^\mu u_\mu = 1$ and
$u^{\mu}=(1,0,0,0)$ in the rest frame.  We use $g^{\mu\nu}=diag\{
1,-1,-1,-1 \}$ as the flat space-time metric.  Besides the
energy-momentum tensor (\ref{T0munu}), a charge current is also
defined and can be expressed in terms of  one proper scalar density
($n$) and the four-velocity field as

\begin{equation}
N_0^{\mu} = n u^{\mu}, \label{eqn:n-ideal} 
\end{equation}
or, equivalently, $n = N_0^\mu u_\mu$, with the four-velocity
normalization given above. 

It should be noted that the proper scalar densities $\varepsilon$, $P$
and $n$ coincide,  respectively, with the energy density, the pressure
and the charge densities  {\it only} in the local rest frame because
of the effect of the Lorentz contraction. 

One can see that the introduced four-velocity field $u^{\mu}$ for an
ideal  fluid satisfies the following equation, 
\begin{equation}
T^{\mu\nu}_0 u_\nu = \varepsilon u^{\mu},
\end{equation}
where $\varepsilon u^{\mu}$ is interpreted as the energy current.
This equation means that $u^{\mu}$ is parallel to the energy current.
On the other hand, from Eq.~(\ref{eqn:n-ideal}), one can see that this
velocity is also parallel to $N^{\mu}_0$.  Therefore, we conclude that
there is no deviation between the energy current and the charge
current in an ideal fluid and, hence, there is no ambiguity for the
definition of the fluid velocity.  However, this situation changes
when the effects of dissipation are taken into account, which is the
case of nonideal fluids.

\subsection{Nonideal fluids}

By using $T^{\mu\nu}_0$ and $N^\mu_0$, which were introduced above,
the general energy-momentum tensor and conserved charge current, in
the presence of dissipative effects, are changed, respectively, to
 
\begin{eqnarray}
T^{\mu\nu}  &=& T_0^{\mu\nu} + \Delta T^{\mu\nu} \;, \label{t1}
\\   N^\mu  &=& N^\mu_0 + \Delta N^\mu \;, 
\label{n1}
\end{eqnarray}
where

\begin{eqnarray}
 \Delta T^{\mu\nu} &=& - (g^{\mu\nu} - u^{\mu} u^{\nu} )\Pi + h^\mu
 u^\nu +  h^\nu u^\mu + \pi^{\mu\nu}\,, \label{Dt1} \\  \Delta
 N^\mu\ &=&  \nu^\mu , 
\label{Dn1}
\end{eqnarray}
and where $\Pi$ is the bulk viscous scalar pressure, $\pi^{\mu\nu}$ is
the shear viscous tensor, $h^\mu$ is the heat current and $\nu^\mu$ is
the diffusion current.  These quantities satisfy the following
orthogonal conditions,

\begin{eqnarray}
&& u^\mu h_\mu = 0 \,, \label{orth_h} \\  && u^\mu \nu_\mu = 0
  \,, \label{orth_nu} \\  && u_\mu \pi^{\mu \nu} = 0\,. 
\label{orth_pi}
\end{eqnarray}
In addition, the shear viscous tensor is traceless, $\pi^{\mu}_\mu
=0$. These dissipative quantities will be explicitly defined below, in
Section~\ref{sec:leading_order}. 

The four new variables, $\Pi,\,h_\mu,\,\nu_\mu$ and $\pi_{\mu\nu}$,
are  introduced to represent the dissipative effects.  However, we can
reduce this number of variables from four to three under an
appropriate choice for the four-velocity in the context of nonideal
fluids.

\subsubsection{Nonideal fluids in the Landau-Lifshitz frame}

The Landau-Lifshitz fluid velocity is defined to satisfy the following
condition, 

\begin{equation}
T^{\mu\nu}u_{\nu} = \varepsilon u^{\mu}. \label{eqn:def-landau}
\end{equation}
Substituting the general expression of the energy momentum tensor
(\ref{t1})  into Eq.~(\ref{eqn:def-landau}), we obtain 

\begin{equation}
T^{\mu\nu}u_{\nu} = \varepsilon u^{\mu} + h^\mu .
\end{equation}
Thus, in the Landau-Lifshitz definition of fluid velocity, $h^\mu =0$.

In short, the energy-momentum tensor and the conserved charge current
in the Landau-Lifshitz theory are then given by

\begin{eqnarray}
T^{\mu\nu} \! &=& \! (\varepsilon_L \!+\! P_L \!+\! \Pi_L)u^{\mu}_L u^{\nu}_L -
g^{\mu\nu}(P_L + \Pi_L)  \nonumber \\
&+& \pi^{\mu\nu}_L,
\label{eqn:tll} 
\\ N^{\mu} &=& n_L u^{\mu}_L + \nu^\mu \;.
\label{eqn:nl1}
\end{eqnarray}
Here the index $L$ indicates the quantities defined in the
Landau-Lifshitz  theory.  {}From the latter equation, one can also
notice that $u^{\mu}_L$ is not parallel  to $N^{\mu}_L$ due to the
diffusion current $\nu^\mu$. This will bring to another possibility of
choice for the fluid four-velocity.

\subsubsection{Nonideal fluids in the Eckart frame}

In the Eckart frame the velocity is defined to satisfy the condition

\begin{equation}
N^{\mu} = n u^{\mu}. 
\label{NuE}
\end{equation}
Substituting the general expression of the conserved charge current
(\ref{n1})  in Eq.~(\ref{NuE}), we obtain 

\begin{equation}
\nu^\mu = 0. 
\end{equation}

Likewise, the energy-momentum tensor and the conserved charge current
in the Eckart theory are defined as
 
\begin{eqnarray}
T^{\mu\nu} &=& (\varepsilon_E + P_E + \Pi_E)u^{\mu}_E u^{\nu}_E -
g^{\mu\nu}(P_E + \Pi_E)  \nonumber \\ &+& h^{\mu}u^{\nu}_E +
h^{\nu}u^{\mu}_E + \pi^{\mu\nu}_E, 
\label{TmunuE}
\\ N^{\mu} &=& n_E u^{\mu}_E .
\label{NmuE}
\end{eqnarray}
Here the index $E$ is used to indicate the quantities defined in the
Eckart frame.

Evaluating $T^{\mu\nu} (u_{E})_\nu$ with the help of the orthogonality
conditions for $h^\mu$ (\ref{orth_h}) and $\pi^{\mu\nu}$
(\ref{orth_pi}), we obtain  

\begin{eqnarray}
T^{\mu\nu} (u_{E})_\nu &=& \varepsilon_E u^{\mu}_E  + h^{\mu}. 
\end{eqnarray}
{}From the above equation, one can notice that $u^{\mu}_E$  is not
parallel to $T^{\mu\nu} (u_{E})_\nu$ due to the heat current $h^\mu$.

\subsection{Connecting the Landau-Lifshitz and Eckart definitions of 
fluid velocity}

In order to discuss the nonrelativistic limit of relativistic
hydrodynamics, it is necessary to express the relativistic
hydrodynamic variables in terms of those defined in the NFS theory.
As a first example, following Landau and Lifshitz~\cite{landau},  the
energy density is given by the proper scalar density  $\varepsilon_L$
expressed in terms of the mass density  $\rho_m$  and the internal
energy per unit mass $\hat{u}$ as 

\begin{eqnarray}
\varepsilon_L = \frac{\rho_m}{\gamma_L} (c^2 + \hat{u}).
\label{eqn:e_L}
\end{eqnarray}

As concerning the expression for the conserved charge density $n_L$,
it is much less trivial.  In nonrelativistic case, the corresponding
conserved charge density is  defined by the number of charged particle
per unit volume.  In the Eckart frame, it is trivial to show that $n_E$ can be
expressed  in terms of the nonrelativistic conserved charge density
as

\begin{equation}
u^\mu_E N_\mu \equiv n_E =
\frac{q}{m}\frac{\rho_m}{\gamma_E} \label{eqn:n_Lu}.
\end{equation}
Note that in the case where the nonrelativistic limit is permitted, the
contribution from  the anti-particle is negligibly small.  On the
other hand, $n_L$ is the proper scalar density associated with
$u^\mu_L$ and  it is clear that 

\begin{equation}
n_L = u^\mu_L N_\mu \neq u^\mu_E N_\mu = n_E.
\end{equation}
That is, to express $n_L$ with $\rho_m$, we need to know the relation
between $n_L$ and $n_E$. 

Because $N^\mu$ can be expressed in the two different ways with
$u^\mu_L$ and  $u^\mu_E$, one can easily find that 

\begin{equation}
n_L = \sqrt{n^2_E - \nu^\mu \nu_\mu}. \label{nenl}
\end{equation}
Substituting Eq. (\ref{eqn:n_Lu}) on the right hand side, we can
express $n_L$ in terms of $\rho_m$.

In short, to introduce nonrelativistic hydrodynamic variables
$\hat{u}$ and $\rho_m$ simultaneously, we need to consider, e.g., both
the Eckart and the Landau rest frames concomitantly.  As was mentioned
in the introduction, this is the reason why the difference of the two
fluid velocities  affects the NLO relativistic correction terms.  It
should be mentioned that this issue has not been discussed in
Refs.~\cite{chandrasekhar,greenberg}.


\section{Fluid equations in the nonrelativistic limit}
\label{sec:leading_order}

Let us now discuss the nonrelativistic limit for the hydrodynamic
equations.  As the first step to obtain the nonrelativistic limit of
the hydrodynamic  variables, it is necessary to specify the
irreversible variables.

\subsection{Nonrelativistic limit of hydrodynamic variables}

We obtain the relativistic covariant expression of the NFS theory when
the linear irreversible thermodynamics (LIT) is applied to determine
the irreversible currents, satisfying the positivity of the entropy
production rate.  However, as was pointed out in Ref.~\cite{dkkm},
such a theory is inconsistent  with the relativistic kinematics in the
sense that the stability of the relativistic fluid changes depending
on the choice of reference frames.  There are several proposals to
calculate these currents, but there is still no established model
(see, for example, Ref.~\cite{dkkm2} for references and discussions
regarding this issue).  However, in the present argument, our intention is
to discuss the behavior of relativistic hydrodynamics in the
nonrelativistic limit and, therefore, the inconsistency mentioned
above will not be of relevance.  Thus, using the following results
obtained in LIT~\cite{landau,mazur}, the linear expressions for the
irreversible variables can be expressed as

\begin{eqnarray}
\nu^{\mu} &=& \frac{\kappa}{c} \left( \frac{n_L T_L}{\varepsilon_L +
  P_L} \right)^2 \Delta^{\mu\nu}_L \partial_\nu
\frac{\mu^{L}_{rel}}{T_L}\,,
\label{nu_landau_1}  
\\ \pi^{\mu\nu}_L &=&  2\,c\, \eta \,\Delta^{\mu\nu\alpha\beta}_L
\partial_\alpha (u_L)_{\beta}   \,,
\label{pi_landau}
\\ \Pi_L &=&  -c \, \zeta\, \partial_\mu u^{\mu}_L ,
\label{Pi_landau}
\end{eqnarray}
where $ \Delta^{\mu\nu}_L$ and $\Delta^{\mu\nu\alpha\beta}_L$ are
projection operators defined as 

\begin{eqnarray}
\Delta^{\mu\nu}_L  &=& g^{\mu\nu} - u^{\mu}_L u^{\nu}_L \,
,\\ \Delta^{\mu\nu\alpha\beta}_L  &=& \frac{1}{2}
(\Delta^{\mu\alpha}_L \Delta^{\nu\beta}_L +
\Delta^{\mu\beta}_L\Delta^{\nu\alpha}_L) -
\frac{1}{3}\Delta^{\mu\nu}_L\Delta^{\alpha\beta}_L\;. 
\end{eqnarray}
Here, $T_L$ and $\mu^L_{rel}$ are the temperature and the chemical
potential, which are obtained by employing the local equilibrium in
the Landau-Lifshitz frame. Let us recall that the chemical potentials
in relativistic and nonrelativistic systems, $\mu^{L}_{rel}$ and
$\mu^{L}_{nrel}$, respectively, are related through $\mu^{L}_{rel} =
mc^2 + \mu^{L}_{nrel}$.  The other coefficients, $\kappa$, $\eta$ and
$\zeta$, appearing in Eqs.~(\ref{nu_landau_1}), (\ref{pi_landau}) and
(\ref{Pi_landau}), represent the coefficients of  the thermal
conductivity, the shear viscosity and the bulk viscosity,
respectively. 

In order to obtain the nonrelativistic limit of relativistic
hydrodynamics, we need to perform an expansion of the hydrodynamic
variables in powers of $v_L/c$, which is a velocity expansion.  For
the linear irreversible variables, we find that the  leading order
contributions are

\begin{eqnarray}
&& \nu^i \propto {\cal O}(v_L^3/c^3)\,, 
\label{nu_i_order}
\\ && \pi^{ij}_L \propto {\cal O}(v_L^0/c^0)\,,
\label{pi_ij_order}
\\ && \Pi_L \propto {\cal O}(v_L^0/c^0).
\end{eqnarray}
Likewise, for the other components, we have that
(see also the argument in Ref.~\cite{landau})

\begin{eqnarray}
&&\nu^0  \propto {\cal O}(v_L^4/c^4)\,,  \\  &&\pi^{00}_L \propto
  {\cal O}(v_L^2/c^2)\,,   \\  &&\pi^{0i}_L = \pi_L^{i0} \propto {\cal
    O}(v_L/c)\,.
\end{eqnarray}
It can be noted that only the purely spatial components of the
irreversible variables,  Eqs.~(\ref{nu_i_order}) and
(\ref{pi_ij_order}), are important at the leading order in a velocity
expansion, whereas the other components contribute only at higher
orders.

By using the velocity expansion, one can obtain the expression of
$n_L$  from Eqs.~(\ref{eqn:n_Lu}) and (\ref{nenl}) as

\begin{equation}
n_L = \frac{q}{m}\rho_m - \frac{q}{2m}\rho_m \frac{{\bf v}^2_L}{c^2} -
\frac{1}{c^4}\frac{q}{8 m}\rho_m {\bf v}^4_L  + {\cal
  O}(v_L^{6}/c^{6})\,.
\label{eqn:n_L}
\end{equation}

Substituting Eq.~(\ref{eqn:n_L}) into Eq.~(\ref{nenl}),  the
fundamental relation between the two fluid velocities in the
nonrelativistic limit is derived as 

\begin{equation}
{v}^{i}_E - { v}^{i}_L = \frac{m c}{q \rho_m}{ \nu}^i + {\cal
  O}(v_L^{4}/c^{4})\,, 
\label{diff_vel}
\end{equation}
where the diffusion current $\nu^i$ can be obtained from
Eq.~(\ref{nu_landau_1}).  Since $\nu^i \propto {\cal
  O}(v_L^{3}/c^{3})$, the two fluid velocities differ only at second
order in the relativistic  corrections, i.e., ${v}^{i}_E -{v}^{i}_L
\propto {\cal O}(v_L^{2}/c^{2})$.

The proper scalar energy density $\varepsilon_L$, when expanded in
$v_L/c$, gives

\begin{eqnarray}
\varepsilon_L &=& \rho_m (c^2 + \hat{u})  + \frac{1}{2}\rho_m {\bf
  v}^2_L  \nonumber \\  &-& \!\!\frac{1}{c^2} \left[  \frac{1}{2}\rho_m
  {\bf v}^2_L \hat{u}  + \frac{1}{8}\rho_m ({\bf v}^2_L )^2 \right]
\!+\!  {\cal O}(v_L^{4}/c^{4}).
\end{eqnarray}
Note that the above equation,  defined in the Landau-Lifshitz frame 
and derived from Eq.~(\ref{eqn:e_L}), gives only part of the relativistic 
corrections to the energy defined from $T^{00}$.

\subsection{Leading order truncation and Navier-Fourier-Stokes theory}

The conservation of energy, momentum and charge are expressed by the
equations of continuity of the energy-momentum tensor and the
conserved charge current,

\begin{eqnarray}
&&\partial_\mu T^{\mu\nu}_L = 0\,,  
\label{eqn:divT} 
\\ &&\partial_\mu N^\mu_L = 0\,.
\label{eqn:divN} 
\end{eqnarray}

By using Eqs.~(\ref{eqn:tll}) and (\ref{eqn:nl1}), we can obtain the
relativistic hydrodynamic model of Landau and Lifshitz, the
Landau-Lifshitz theory.  The nonrelativistic limit of this theory can
be obtained from the substitution of the relativistic hydrodynamic
variables by the leading order expressions for these variables that we
have obtained in the previous section.  In the present work, we adapt
the following orders for the hydrodynamic variables  $\hat{u}$, $P_L$
and $\mu^L_{nrel}$, 

\begin{eqnarray}
\frac{\hat{u}}{c^2},\  \frac{P_L}{c^2},\  \frac{\mu^L_{nrel}}{c^2}
\propto {\cal O}(v_L^2/c^2).
\label{nonrel_cond}
\end{eqnarray}
Then, as we will see soon later, the NSF theory is reproduced in the
leading order approximation.

As it was shown in Eq.~(\ref{diff_vel}), the difference of the two
fluid velocities, ${\bf v}_E$ and ${\bf v}_L$, appears only at order
${\cal O}(v_L^{2}/c^{2})$.  By truncating at ${\cal O}(v_L^{0}/c^{0})$
of the velocity expansion, we simply have that

\begin{equation}
{\bf v}_{E} = {\bf v}_L = {\bf v}\,.
\label{eq:vvv}
\end{equation}
This equality shows that the velocities  ${\bf v}_L$ and ${\bf v}_{E}$
define the same rest frame at the leading order in the $v_L/c$
expansion.  Thus, the energy and mass flows are both parallel to the
fluid velocity, which is the case of the usual nonrelativistic
hydrodynamics. 

The hydrodynamic variables $P_L$, $\Pi_L$ and $\pi_L^{ij}$ occurring
in the energy-momentum tensor $T^{\mu\nu}_L$ are obtained by employing
the local equilibrium in the Landau-Lifshitz frame.  It should be
noted, however, that  there is a unique rest frame because of
Eq.~(\ref{eq:vvv}) and these hydrodynamic variables do not have any
frame dependences in the leading order truncation.  Therefore, we can
verify that the nonrelativistic limit of relativistic hydrodynamics
reproduces the NFS theory,

\begin{eqnarray}
&& \partial_t \rho_m +  \nabla \cdot \left( \rho_m {\bf v}_L \right)
  =0\,,
 \label{eq:rho_NS} 
\\ && \rho_m \left(\partial_t + {\bf v}_L \cdot \nabla \right) {
  v}^{i}_L = - \sum_{j=1}^3 \nabla_j {\cal P}^{ij}_{L0} \,, 
\label{eq:v_NS}
\\ && \rho_m \left( \partial_t \!+\! {\bf v}_L \cdot \nabla \right)
\hat{u} \!= \!-\! \nabla \cdot {\bf q} \!- \!\!\sum_{j,k=1}^3 {\cal
  P}^{jk}_{L0}\nabla_j { v}^k_L,
\label{eq:u_NS}    
\end{eqnarray}
where the stress tensor ${\cal P}^{ij}_{L0}$ is given by

\begin{eqnarray}
{\cal P}^{ij}_{L0} = \delta^{ij}(P_L + \Pi_L) + { \pi}_L^{ij}
\,, \label{eqn:pl0}
\end{eqnarray}
with the viscosities $\Pi_L$ and $\pi^{ij}_L$, which are also used in
the next Section, are defined by their usual leading order
expressions~\cite{landau}

\begin{eqnarray}
\Pi_L &=& - \zeta   \nabla \cdot {\bf v}_L \,, \label{eqn:PI0}
\\ \pi^{ij}_L &=& - \eta \left( \nabla_i {\bf v}_{Lj}  + \nabla_j {\bf
  v}_{Li} - \frac{2}{3}\delta_{ij}\nabla \cdot {\bf v}_L \right)
\,, \label{eqn:pi0}
\end{eqnarray}
while the heat current vector ${\bf q}$ is defined by

\begin{eqnarray}
 {\bf  q} = -  \kappa \nabla T_L \,.
\label{q_1}
\end{eqnarray}


\section{Next-to-leading order relativistic corrections}
\label{sec4}

To determine the difference between the two fluid velocities,  we
calculate the NLO corrections for the NFS theory.   By keeping the
relativistic corrections terms up to ${\cal O}(v_L^{2}/c^{2})$  in the
fluid equations (\ref{eq:rho_NS}), (\ref{eq:v_NS}) and
(\ref{eq:u_NS}),  we obtain the mass, the momentum and the energy
equations,  respectively, as

\begin{eqnarray}
&&\partial_t \rho_m +  \nabla \cdot \left( \rho_m {\bf v}_L \right) =
  - \frac{mc}{e} \nabla \cdot {\bf \nu} + {\cal O}(v_L^{4}/c^{4}), 
\label{nlo_LL_number}
\end{eqnarray}

\begin{eqnarray}
&&\rho_m\left(\partial_t + {\bf v}_L \cdot \nabla \right){ v}^{i}_L =
  \nonumber \\ && - \sum_{j=1}^3 \left[ \delta_{ij}  -
    \frac{\delta_{ij}}{c^2}\left(\frac{{\bf v}_L^2 + 2\hat{u}}{2}  +
    \frac{P_L + \Pi_L}{\rho_m}\right)  - \frac{1}{c^2} \frac{ {
        \pi}^{ij}_L}{\rho_m} \right] \nonumber \\ && \times \nabla_{j}
  (P_L + \Pi_L)   \nonumber \\ &&-  \sum_{j,k=1}^{3} \left[
    \delta_{ij} - \frac{\delta_{ij}}{c^2}\left(\frac{{\bf v}_L^2 +
      2\hat{u}}{2} + \frac{P_L + \Pi_L}{\rho_m}\right)
    \right. \nonumber \\ && \left. - \frac{1}{c^2}\left({ v}_L^i {
      v}_{L}^j +  \frac{ { \pi}^{ij}_L}{\rho_m}\right) \right]\nabla_k
            { \pi}^{jk}_L  \nonumber \\ &+& \frac{1}{c^2}
            \sum_{j,k=1}^3 \left({ v}_L^i  { \pi}^{jk}_L +  {
              v}_L^k\ { \pi}^{ij}_L \right)\nabla_k { v}_L^j  -
            \frac{1}{c^2} { v}_L^i \partial_t ( P_L + \Pi_L )
            \nonumber \\  &-& \frac{1}{c^2} \sum_{j=1}^3  { v}_{L}^j
            \partial_t { \pi}^{ij}_L +  {\cal O}(v_L^{4}/c^{4}), 
\label{nlo_LL_momentum}
\end{eqnarray}

\begin{eqnarray}
&&\rho_m\left( \partial_t + {\bf v}_L \cdot \nabla \right)\hat{u}
  \nonumber \\ &&= - \sum_{i,j=1}^3 \left\{\left( 1 +
  \frac{1}{2}\frac{{\bf v}_L^2}{c^2}\right)\left[\delta_{ij}(P_L +
    \Pi_L) + { \pi}^{ij}_L\right]   \right. \nonumber \\ && \left. -
  \frac{1}{c^2}\sum_{k=1}^3 {v}_{L}^i { v}_L^k{ \pi}^{jk}_L \right\}
  \nabla_i { v}_L^j \nonumber \\ && +   \frac{1}{c^2} \sum_{i=1}^3
  \left({ v}_{L}^i \frac{P_L + \Pi_L}{\rho_m} +  \sum_{k=1}^3 {
    v}_{L}^k \frac{{ \pi}^{ik}_L}{\rho_m}\right)  \nonumber \\ &&
  \times \left[\nabla_i (P_L + \Pi_L) + \sum_{j=1}^3 \nabla_j{
      \pi}^{ij}_L\right] \nonumber \\ && + \frac{mc^3}{q}\sum_{i=1}^3
  \nabla_i\left( { \nu}^i -  \sum_{j=1}^3 \frac{{ v}_L^i}{c} \frac{{
      v}_L^j}{c} { \nu}^j\right)   \nonumber \\ && + \frac{mc}{q}
  \hat{u} \nabla \cdot {\bf \nu}  +  {\cal O}(v_L^{4}/c^{4}).
\label{nlo_LL_energy}
\end{eqnarray}
The irreversible variables, which are also expanded up to the
next-to-leading order ${\cal O}(v_L^{2}/c^{2})$, are given by

\begin{eqnarray}
\nu^{i} &=& \kappa \frac{q}{m c^3 } \left[ \left(\nabla_i-
  \frac{\hat{u}}{c^2}\nabla_i  + \frac{v^{i}_L}{c^2}D_L \right) T_L
  \right. \nonumber \\ &-& \left.  \frac{T_L}{\rho_m c^2} \nabla_i
  P_L  \right]\,,
\label{nu_landau}  
\end{eqnarray}

\begin{eqnarray}
\pi^{ij}_L&=& -\eta  \left\{  \partial_j{\bf v}_{Li}+\partial_i{\bf
  v}_{Lj}  +
\frac{1}{c^2}\left[\left({\bf v}_{Li}\partial_j + {\bf
    v}_{Lj}\partial_i \right) \left(\frac{{\bf v}_{L}^2}{2}\right)  
\right.  \right. \nonumber \\ &+&\! \left. \left.
  \frac{{\bf v}_{L}^2}{2} \left(\partial_j{\bf v}_{Li} \!+\! \partial_i{\bf
    v}_{Lj}\right)   \!-\!
       {\bf v}_{Lj} D_L {\bf v}_{Li}  - {\bf v}_{Li} D_L {\bf v}_{Lj}
       \right]\right\} \nonumber\\ &+&\frac{2}{3}\eta \left\{\nabla
\cdot {\bf v}_{L} \delta_{ij} +\frac{1}{c^2} \left[\left(\frac{{\bf
      v}_{L}^2}{2} \delta_{ij} + {\bf v}_{Li}{\bf v}_{Lj}\right)
  \nabla \cdot {\bf v}_{L} \right. \right. \nonumber \\ &+&
  \left. \left.  D_L \left(\frac{{\bf v}_{L}^2}{2}\right) \delta_{ij}
  \right]\right\},
\label{pi_landau2}
\end{eqnarray}

\begin{eqnarray}
\Pi_L \!&=&\!  - \!\zeta\, \left\{\nabla\cdot {\bf v}_L \!+\!
\frac{1}{c^2}\left[D_L\left(\frac{{\bf v}_L^2}{2}\right) \!+\!
  \frac{{\bf v}_L^2}{2}\nabla\cdot {\bf v}_L\right] \right\}.
\label{Pi_landau2}
\end{eqnarray}

One can notice that in order to satisfy the NLO energy and momentum
conservation equations,  the relativistic correction terms appearing
in the definitions of energy and momentum should be considered.  The
expressions for the nonrelativistic energy density and momentum
current, $\rho_m \hat{e} \equiv \rho_m(\frac{1}{2}{\bf
  v}^2+\hat{u})$ and $\rho_m{\bf m} \equiv \rho_m{\bf v}$,
respectively, which are conserved in the NFS theory, are no longer
conserved in the NLO equations.  However, the appropriate expressions
in the relativistic context are exactly those obtained from the
components of the relativistic energy-momentum tensor,  $\rho_m
\hat{e}_L = T^{00}$  and $\rho_m {\bf m}^i_L = T^{0i}/c$,
respectively. Thus, from the energy-momentum tensor in the
Landau-Lifshitz frame, Eq.~(\ref{eqn:tll}), the expressions for the
energy per unit mass and  for the momentum per unit mass, that account
up to the ${\cal O}(v_L^{2}/c^{2})$ relativistic corrections, are
given, respectively,  by

\begin{eqnarray}
\hat{e}_L  &=& c^2 + \left(\frac{{\bf v}_L^2}{2} + \hat{u}\right) +
\frac{1}{c^2}\left[\left(\frac{3}{8}{\bf v}_L^2 + \frac{1}{2}\hat{u}
  \right){\bf v}_L^2 \right. \nonumber \\ &+&
  \left. \sum_{i,j=1}^3\frac{(P_L + \Pi_L)\delta^{ij} +{
      \pi}^{ij}_L}{\rho_m} { v}_L^i { v}_L^j\right],
\label{hateL}
\end{eqnarray}
and

\begin{eqnarray}
{m}^{i}_L  &=& { v}_L^i +  \frac{1}{c^2} \left[ \left( \frac{1}{2}
  {\bf v}_L^2 +   \hat{u} \right){ v}_L^i     \right.  \nonumber
  \\ &+& \left. \sum_{j=1}^3 \frac{(P_L + \Pi_L) \delta^{ij} +  {
      \pi}^{ij}_L}{\rho_m}{ v}_L^j \right].
\label{miL}
\end{eqnarray}
{}From these expressions, Eqs.~(\ref{nlo_LL_number}),
(\ref{nlo_LL_momentum}) and (\ref{nlo_LL_energy}) can be cast into
much simpler forms,

\begin{eqnarray}
&&\partial_t \rho_m +  \nabla \cdot \left( \rho_m {\bf v}_E \right)  =
0, 
\label{nlo_LL2_number}
\\ &&\rho_m ( \partial_t +{\bf v}_E \cdot \nabla ){ m}^{i}_L   =  -
(\nabla \cdot {\cal P}_L)^i , 
\label{nlo_LL2_momentum}
\\
&& \!\!\!\!\!\!\!\!\!\rho_m \left( \partial_t \!+\! {\bf v}_E  \cdot \nabla \right)\hat{e}_L
= \!-\! \nabla \cdot {\bf j}_L \!-\! \sum_{i=1}^3 \partial_i  (  {\cal P}_L
\cdot {\bf v}_L)^i,
\label{nlo_LL2_energy}
\end{eqnarray}
where we have used Eq.~(\ref{diff_vel}) to take ${\bf v}_E$ into
account.  Here, $ (\nabla \cdot {\cal P}_L)^i=  \sum_{j=1}^3
\partial_j {\cal P}^{ij}_L$ and $ ( {\cal P}_L \cdot {\bf v}_L )^i =
\sum_{j=1}^3 {\cal P}^{ij}_L { v}^j_L$.  The spatial components for
the stress tensor ${\cal P}_L$
and for the heat current vector ${\bf j}_L$ are, respectively, given by

\begin{eqnarray}
{\cal P}^{ij}_L  &=&  {\cal P}^{ij}_{L0}  - \frac{1}{c^2}\sum_{k=1}^3{
  \pi}^{ik}_L { v}_L^k { v}_L^j  - \frac{m c}{q} { m}^{i}_L {
  \nu}^{j}\, , 
\label{PijL}
\end{eqnarray}
and 

\begin{eqnarray}
\!\!\!\!\!\!{ j}^{i}_L  &=& - \frac{m c}{q} \left( \hat{e}_L \, { \nu}^{i}  -
\sum_{j=1}^3{ m}^{i}_L  \, { \nu}^{j}{ v}_L^j\right) 
\nonumber \\ 
&=& \! -\!  \kappa \left[ \left( \nabla_i \!+\! \frac{{\bf v}^2_L}{c^2} \nabla_i \!+\!
  \frac{v^{i}_L}{c^2}  \partial_t \right) T_L  \!-\! \frac{T_L}{\rho_m
    c^2} \nabla_i  P_L \right],
\label{jiL}
\end{eqnarray}
where, in the calculation of ${ j}^{i}_L $, we have used the explicit
expression for $\nu^i$ given by Eq.~(\ref{nu_landau}). One can easily
check that the relativistic quantities $\rho_m \hat{e}_L$ and $\rho_m
{\bf m}_L$ are conserved densities in the next-to-leading order
relativistic hydrodynamics, as well as $\rho_m \hat{e}$ and $\rho_m
{\bf m}$ are in the NFS theory.

The hydrodynamic equations given by Eqs.~(\ref{nlo_LL2_number}),
(\ref{nlo_LL2_momentum}) and (\ref{nlo_LL2_energy}), represent the
main result of this work.  Most importantly, we note that the two
fluid velocities ${\bf v}_E$ and ${\bf v}_L$ appear in the above
equations.  This shows that it is possible to work concurrently with
two different velocities in the NLO hydrodynamics.  Of course, one of
the velocities is eliminated by  substituting the relation given by
Eq. (\ref{diff_vel}), but the  above expressions are essential  to
compare with bivelocity hydrodynamics in the next section (see also
the comment below Eq.~(\ref{diff_vel_bi})).  We also notice that the
last term on the right-hand side of the energy equation,
Eq.~(\ref{nlo_LL2_energy}), associated with the work done by the
stress tensor, is implemented following ${\bf v}_{L}$.  On the other
hand, in bivelocity hydrodynamics, this term follows ${\bf v}_{V}$.
The discussion concerning this term plays an important role in the
comparison.

Before closing this section, as a final remark concerning the 
stress tensor defined by Eq.~(\ref{PijL}), we note that it is asymmetric.
But this is only because we wrote Eq.~(\ref{nlo_LL2_momentum}) in a way
it can be compared to the bivelocity hydrodynamics in the next section.
In fact, the momentum equation  in the NLO, Eq.~(\ref{nlo_LL2_momentum}),
can be expressed with a symmetric tensor as 

\begin{eqnarray}
\partial (\rho_m m^{i}_L) 
= - \sum_j \partial_j \tilde{\cal P}^{ij}_L,
\label{eq:Psymm}
\end{eqnarray}
where, when we move the term depending on the Eckart
fluid-velocity in Eq.~(\ref{nlo_LL2_momentum}) to the right-hand-side of 
that equation, and upon using also Eq.~(\ref{nlo_LL2_number}), 
we obtain that

\begin{eqnarray}
\!\!\!\!\tilde{\cal P}^{ij}_L &=& P^{ij}_{L0} \nonumber \\
&+&  \!\!\frac{1}{c^2} \left( \rho_m c^2 \!+\! \frac{\rho_m}{2}{\bf v}^2_L \!+\! 
\hat{u} \!+\! P_L \!+\! \Pi_L \right){\bf v}^{i}_L {\bf v}^{j}_L, 
\label{Psymm}
\end{eqnarray}
which is symmetric.


\section{Comparison with Bivelocity Hydrodynamics}
\label{sec6}

As it was shown in the previous section, the NLO relativistic
corrections to the NFS theory leads to a new hydrodynamic model that
is described by the two different fluid velocities, ${\bf v}_{E}$ and
${\bf v}_L$.  In this section, we compare this NLO equations with
bivelocity hydrodynamics.

\subsection{Velocity in Landau-Lifshitz frame and volume velocity}

Bivelocity hydrodynamics is constructed with the mass velocity ${\bf
  v}_M$ and the volume velocity ${\bf v}_V$.  The mass velocity is
defined to be parallel to the mass flow as usual.   The origin of the
volume velocity is attributed to the fact that the flow of the
constituent particles of the fluid (velocity of the tracer particles)
is not necessarily parallel to the mass velocity
\cite{brenner,brenner2,brenner2012,brenner2012-2,brenner2013}.

The velocity ${\bf v}_{E}$ in relativistic hydrodynamics is defined to
be parallel to the conserved charge current and, therefore, it is
quite  natural to ask if it can be identified with the mass
velocity. However, it is not trivial to know in principle whether
${\bf v}_L$ corresponds to the volume velocity, because the physical
meaning of these two velocities seem to be different.  Thus, we need
to investigate the explicit relation between ${\bf v}_{L}$ and ${\bf
  v}_V$.

The relation between ${\bf v}_M$ and ${\bf v}_V$ is known in the
context of bivelocity hydrodynamics and is given by~\cite{brenner}

\begin{eqnarray}
\lefteqn{{\bf v}_{M} - {\bf v}_V = - C_v \frac{\eta}{\rho_m}  \nabla
  \ln \rho_m } \nonumber \\  &=&  - C_v \frac{\eta}{\rho^2_m}
\left\{ \left( \frac{\partial \rho_m}{\partial T}\right)_P \nabla T +
\left( \frac{\partial \rho_m}{\partial P}\right)_T \nabla P
\right\}\,, 
\label{diff_vel_bi}
\end{eqnarray}
where the coefficient $C_v$ is a free parameter that  can be obtained
once a particular application or theory is given.  {}For example, some
results for the coefficient $C_v$  can be found in Table I of
Ref.~\cite{brenner2012}.

It is clear from Eq.~(\ref{diff_vel_bi}) that the volume velocity is
expressed in terms of the mass velocity and, thus, these are not
independent.  This is simply the notation introduced in bivelocity
hydrodynamics, and for the sake of the comparison with this theory, we
will also express our results of the previous section by using the two
velocities.  In this manner, although the existence of the two fluid
velocities is assumed in bivelocity hydrodynamics, they do not
represent independent variables.

On the other hand, as it was shown in Eq.~(\ref{diff_vel}),  the
difference between the two fluid velocities in relativistic
hydrodynamics  is given by 

\begin{eqnarray}
 {\bf v}_{E} - {\bf v}_L  &=& \frac{mc}{q\rho_m} \nu^{i}  \nonumber
 \\ &=& \kappa \frac{1}{c^2 \rho_m} \left[ \left(\nabla_i-
   \frac{\hat{u}}{c^2}\nabla_i  + \frac{v^{i}_L}{c^2}D_L \right) T_L
   \right. \nonumber \\ &-& \left.  \frac{T_L}{\rho_m c^2} \nabla_i
   P_L  \right]\,,
\label{diff_vel_ll}
\end{eqnarray}
where $D_L = \partial_t + {\bf v}_L \cdot \nabla$ and we  have used
the explicit expression of $\nu^i$, Eq.~(\ref{nu_landau}).  Here we
have expanded $\nu^i$ using the thermodynamic relation, 

\begin{equation}
\nabla_i \frac{\mu^L_{rel}}{T_L} = - \left( \frac{\varepsilon_L +
  P_L}{n_L T^2_L} \right)\nabla_i T_L  + \left( \frac{1}{n_L T_L}
\right) \nabla_i P_L .
\end{equation}

It can be verified that $D_L T_L$ is approximately given by $\nabla^2
T_L$  when the contribution from the energy dissipation is
sufficiently  small in the energy equation.  Then, the third term in
the right hand side of Eq.~(\ref{diff_vel_ll}) is a higher-order
contribution of the spatial derivative and can be neglected.  The
difference between ${\bf v}_{E}$ and ${\bf v}_{L}$  is, then,
determined by the gradients of temperature and pressure similarly to
the case of ${\bf v}_M$ and ${\bf v}_V$.  Therefore, it can be
concluded that the volume velocity in bivelocity hydrodynamics ${\bf
  v}_V$ is related to the velocity in the Landau-Lifshitz frame ${\bf
  v}_L$.

\subsection{Equations in bivelocity hydrodynamics}

Let us investigate further whether the NLO equations given by
Eqs.~(\ref{nlo_LL2_number}), (\ref{nlo_LL2_momentum}) and
(\ref{nlo_LL2_energy}) can have the same structure as the fluid
equations in bivelocity hydrodynamics.

In the following,  we use ${\bf v}_V = {\bf v}_L$ and ${\bf v}_M =
{\bf v}_{E}$  to express the equations in bivelocity hydrodynamics  to
avoid confusion in the comparison.

The bivelocity hydrodynamics model is characterized by the following
set of equations~\cite{brenner2012,brenner2012-2}, 

\begin{eqnarray}
&& \partial_t \rho_m +  \nabla \cdot \left( \rho_m {\bf v}_{E} \right)
  = 0, 
\label{bivelo1}
\\ && \rho_m \left(\partial_t + {\bf v}_{E} \cdot \nabla \right)
   {m}^{i}_{bi} =  - (\nabla \cdot {\cal P}_{L0})^i , 
\label{bivelo2}
\\ && \rho_m (\partial_t + {\bf v}_{E} \cdot \nabla) \hat{e}_{bi} =  -
\nabla \cdot {\bf j}_u -  \nabla \cdot ( {\cal P}_{L0} {\bf v}_{L}),
\label{bivelo3}    
\end{eqnarray}
where 

\begin{eqnarray}
&&{\bf m}_{bi} = {\bf v}_{E} , \label{eqn:mb} \\  &&\hat{e}_{bi}  =
  \frac{{\bf v}_{E}^2}{2} + \hat{u} \,. 
\label{eqn:eb}
\end{eqnarray}
On the other hand, the heat current vector in bivelocity
hydrodynamics,  by using LIT, is given by 

\begin{eqnarray}
{\bf j}_u &=& -\left[ \kappa + \frac{C_v \eta }{\rho^2_m} \left(
  \frac{\partial \rho_m}{\partial T} \right)_P \right] \nabla T
\nonumber \\ &+& \frac{C_v \eta}{\rho^2_m} \left[ T \left(
  \frac{\partial \rho_m}{\partial T} \right)_P  - P  \left(
  \frac{\partial \rho_m}{\partial P} \right)_T  \right] \nabla P.
\label{j_u}
\end{eqnarray}

One can note that the heat current is given by the linear combination
of the two  thermodynamic forces: One is for the pure heat conduction,
$\nabla T$; and the other is induced by the existence of  the volume
velocity, $\nabla P$.  Then, because of the Curie principle, the most
general expression is given by their  linear combination.

\subsection{Comparison between the two approaches}

By comparing the NLO equations (\ref{nlo_LL2_number}),
(\ref{nlo_LL2_momentum}) and (\ref{nlo_LL2_energy}) with those of the
bivelocity hydrodynamics, Eqs.~(\ref{bivelo1}), (\ref{bivelo2}) and
(\ref{bivelo3}), one can find that the structures of the two theories
are similar.  In fact, the various assumptions used  in the derivation
of bivelocity hydrodynamics are naturally reproduced in the NLO
equations.  

In both theories the equations are expressed with the material
(substance) derivative for the mass velocity ${\bf v}_{E}$.  That is,
the evolution of the hydrodynamic variables are defined in terms of
the fluid element, which moves with the mass velocity.  However, the
work done by the stress tensor, which appears in the second terms  on
the right hand side of the energy equation of each theory,  and the
forms of the bulk and shear viscosities are given in terms of the
volume velocity ${\bf v}_L$, but not ${\bf v}_{E}$.

It is also verified that in both theories the heat currents are
induced even by the pressure gradient.  In bivelocity hydrodynamics,
this behavior is because LIT leads to the pressure gradient as the
thermodynamic force associated with the volume velocity
\cite{brenner2012,brenner2012-2}.  On the other hand, in the NLO
equations, the thermodynamic force associated with the diffusion
current $\nu^{\mu}$ is given by the gradient $\nabla (\mu/T)$ and the
pressure gradient is induced by the chemical potential dependence
included in this term.  These behaviors are assumed in the derivation
of bivelocity hydrodynamics, while the very same  are automatically
reproduced in the NLO equations.  The consistency that we found in the
comparison above can be considered as an indication of support for the
validity of the application of LIT for the construction of bivelocity
hydrodynamics.  

There are still qualitative differences that we cannot ignore: 1) the
energy and momentum variables definitions in bivelocity hydrodynamics
are given in terms of the mass velocity, which is argued to be the
universal behavior~\cite{brenner2012}, whereas, in the NLO equations,
these variables are defined in terms of the volume velocity, and  2)
the symmetric stress tensor in bivelocity hydrodynamics ${\cal
  P}_{L0}^{ij}$ is replaced by an asymmetric  one in the NLO equations
${\cal P}_{L}^{ij}$.  However, these problems are essentially
connected to, and can be explained as an effect of the relativistic
corrections, which are not considered in the original formulation of
bivelocity hydrodynamics.

In the framework of bivelocity hydrodynamics, in fact,  the
definitions of energy and momentum are not trivial because of the
existence of the two different fluid velocities. Then, in
Ref.~\cite{brenner2012}, it is assumed that the nonrelativistic forms
of the energy and momentum are still  held even in the formulation of
bivelocity hydrodynamics. In other words, ${\bf m}_{bi}$ and
$\hat{e}_{bi}$  are, respectively, given by the linear and quadratic
functions of a  certain velocity. Under this assumption, Brenner
succeeded in showing that  this velocity is given by the mass
velocity. In short, Eqs. (\ref{eqn:mb})  and (\ref{eqn:eb}) are
derived.  It is however obvious that this argument is not applicable
to the NLO equations  because the definitions of momentum and energy
are modified due to the relativistic corrections. Therefore, we can
still consider the momentum and energy  per unit mass as functions of
the volume velocity in the present case.

The modification of the velocity dependence in the energy and momentum
of  bivelocity hydrodynamics naturally leads to the introduction of an
asymmetric  stress tensor. It is difficult to predict the velocity
dependence when the relativistic  corrections are allowed to be
included in the framework of bivelocity hydrodynamics.  However, as a
simplest example, let us consider the case where the momentum is
simply ${\bf m}_{bi} = {\bf v}_{L}$.  Then, the momentum equation
(\ref{bivelo2}) is expressed as

\begin{equation}
\partial_t ( \rho_m {\bf v}^{i}_L )  = - \sum_{j=1}^3 \nabla_j ({\cal
  P}^{ij}_{L0} +  \rho_m {\bf v}^j_{E} {\bf v}^{i}_L )\;.
\label{eqPl02}
\end{equation}
Note that the second term on the right-hand side in Eq.~(\ref{eqPl02})
has two velocities;  one comes from the definition of ${\bf m}_{bi}$
and the other from the material derivative.  As a consequence, the
second rank tensor on the right-hand side in Eq.~(\ref{eqPl02}),
$({\cal P}^{ij}_{L0} + \rho_m {\bf v}^j_{E} {\bf v}^{i}_L)$,  is {\it
  not}  symmetric for the exchange of the indexes $i$ and $j$ and,
hence, the angular momentum density defined by $\rho_m ({\bf r} \times
{\bf m}^{i}_b) = {\bf r} \times \rho_m {\bf v}^{i}_L$ is not a
conserved density.   Therefore, to satisfy the angular momentum
conservation in this case, the stress tensor  ${\cal P}^{ij}_{L0}$
should contain an asymmetric part that cancels the last term on the
right hand side, $\rho_m {\bf v}^j_{E} {\bf v}^{i}_L$.  This
conclusion is still the same even if we consider a more complex
velocity dependence. In fact, one can easily confirm that the angular 
momentum density $\rho_m ({\bf r} \times
{\bf m}^{i}_L)$ is conserved in the NLO equation (\ref{eq:Psymm}).

In summary, the formulation of bivelocity hydrodynamics is based on
various assumptions.  Most of these assumptions (material derivatives,
work done by the stress tensor, the forms  of the viscosities, the
thermodynamic force of the volume velocity) are automatically
reproduced in the NLO equations.  On the other hand, the stress tensor
in bivelocity hydrodynamics is given by the symmetric form,  while the
corresponding stress tensor in the NLO equations is asymmetric.  That
is, the structure of the NLO equations do not reproduce bivelocity
hydrodynamics completely.  However, this is because bivelocity
hydrodynamics is not constructed in such a way to include the
relativistic corrections.  When we generalize the argument to include
the relativistic corrections, an asymmetric stress tensor  emerges
even in bivelocity hydrodynamics in order to satisfy the angular
momentum conservation.  That is, the nonrelativistic hydrodynamics
with the NLO corrections is qualitatively equivalent to this
generalized version of bivelocity hydrodynamics.


\section{Concluding Remarks}
\label{sec7}

In this work, we have discussed the nonrelativistic limit of
relativistic hydrodynamics.  In relativistic hydrodynamics, it is
possible to define two fluid velocities;  one is parallel to the
energy current and the other can be defined to be parallel  to the
conserved charge current.  The difference between these velocities
disappears in the nonrelativistic limit and the  NFS theory is
reproduced. 

{}From the results we have obtained, we do not observe any bivelocity
effect  in the nonrelativistic limit,  but it does not necessarily
mean that  bivelocity hydrodynamics does not exist in the
nonrelativistic regime.  It is because the relativistic hydrodynamics
used here is obtained  by employing the linear irreversible
thermodynamics  and hence the possible nonlinear effects in the
irreversible currents are not considered.  If such effects are taken
into account, the correction terms may appear even in  this regime.
As a matter of fact, there are arguments  that the bivelocity effect
can be induced from such nonlinearities (see, for example,
Refs.~\cite{dzy,kli,ott,grau,gree,daz,eu,don,koide}).  How these
nonlinearities also manifest in the context of relativistic
hydrodynamics is an interesting subject to be explored in a future
work.

Afterwards, we have explicitly obtained the NLO relativistic
corrections to the NFS theory for the case of the Landau-Lifshitz
definition of fluid velocity. Previous studies of the NLO relativistic
corrections to the NSF theory were available only for a Eckart
fluid~\cite{greenberg}.  We believe that our results represent an
important contribution to the study of cases where the Eckart scenario
does not apply. 

The derived NLO hydrodynamics can be  expressed concurrently in terms
of both fluid velocities, where one of them is a  Eckart fluid
velocity and the other the Landau-Lifshitz fluid velocity.   Comparing
this result with bivelocity hydrodynamics, we found that the
Landau-Lifshitz velocity can be identified with the volume velocity in
bivelocity hydrodynamics.  

Using this identification, we have confirmed that most of the
assumptions used in bivelocity hydrodynamics (material derivatives,
work done  by the stress tensor, the forms of the viscosities, the
thermodynamic force of the volume velocity)  are automatically
reproduced in the NLO equations. The difference comes from the stress
tensor;   the stress tensor in the bivelocity hydrodynamics is
symmetric, while  the one in the NLO equations is asymmetric.
However, this difference can be explained by the different origins of
the volume velocity; in bivelocity hydrodynamics, the volume velocity
is induced as a consequence of the definition of the diffusive flux of
volume ${\bf j}_v$ (which is absent in the NFS theory), while in the
NLO equations it appears as an effect of the relativistic corrections.
Then, by discussing the symmetry properties of the stress tensors  in
connection to the angular momentum conservation  in both theories, we
have found that the argument of bivelocity hydrodynamics can be
extended so as  to include the relativistic corrections, and then the
stress tensor is permitted to be  asymmetric even in bivelocity
hydrodynamics to satisfy the angular momentum conservation.  In short,
in the sense discussed above, the hydrodynamics including the NLO
corrections is qualitatively equivalent to bivelocity hydrodynamics.

In the original idea of bivelocity hydrodynamics, the origin of the
volume velocity is identified with the flow of the constituent
particles of the fluid that is not parallel to the mass velocity, and
this deviation is enhanced for the compressible fluid.  However, as we
have shown in the present work,  a similar situation can be expected
as the result of relativistic effects  and that is possible to be
observed even for incompressible fluids.  The study we have performed
in this work points out, thus, that analogous effects expected from
the bivelocity picture can be obtained by observing the behavior of,
for example, high energy fluids in cosmology and relativistic
heavy-ion collisions. These are in fact areas of research that our
results may have immediate applications and that are worth of future
investigation. 


\acknowledgments

T.K. acknowledge discussions with G. S. Denicol, X. Huang and
L. S. Garc\'{i}a-Colinin  in the initial stages of the present study
and financial support from Conselho Nacional de Desenvolvimento
Cient\'{\i}fico e Tecnol\'ogico (CNPq).  R.O.R. is partially supported
by research grants from CNPq and Funda\c{c}\~ao Carlos Chagas Filho de
Amparo \`a Pesquisa do Estado do Rio de Janeiro (FAPERJ). G.S.V. is
supported by Coordena\c{c}\~ao de Aperfei\c{c}oamento de Pessoal de
N\'{\i}vel Superior (CAPES).


\end{document}